# A Preliminary Study of Multilingual Code Language Models for Code Generation Task Using Translated Benchmarks

Rohit Dandamudi
rohitd@mail.ubc.ca
University of British Columbia
Kelowna, BC, Canada

Gema Rodríguez-Pérez
gerope@mail.ubc.ca
University of British Columbia
Kelowna, BC, Canada

## ABSTRACT

Evaluating the performance of Code Language Models (CLMs) for software engineering tasks, especially in multilingual and low-resource programming language settings, poses significant challenges. These challenges are primarily due to the lack of high-quality benchmarks across various programming languages and the imbalanced nature of the CLMs training corpus. Although recent advances in one of the common downstream tasks, code generation, have shown promise by introducing translated benchmarks using different methodologies, there is a current lack of empirical evidence assessing these benchmarks. To address this gap, we conducted a preliminary study to evaluate the performance of Poly-Coder, a pioneering open-source, multilingual CLM built for code generation. We utilized two existing state-of-the-art translations of the popular code generation benchmark, HumanEval, facilitated by the OctoPack and MultiPL-E studies. Our results suggest that the outcomes observed in these translated benchmarks align well with evaluation metrics used during the training phase, such as perplexity, thereby validating their effectiveness in estimating the performance of CLMs. However, we identified several inconsistencies in the CLMs' performance across the translated benchmarks and encountered challenges in replicating the results. These initial insights highlight the need for more comprehensive empirical studies to fully understand translated benchmarks' methodological approaches, limitations, and reproducibility. Such studies are essential to ensure their reliability before they are widely adopted.

## 1 INTRODUCTION

Significant research is being conducted to enhance the performance of Large Language Models (LLMs) in coding tasks [16]. These tailored models are also known as Large Language Models for Code or Code Large Language Models, and previous research focuses on assessing their performance in different software engineering downstream tasks, such as code generation [4], code summarization [1], code translation [18], and code clone detection [17], among others. We will refer to these models as Code Language Models (CLMs) in this study; these CLMs have been integrated into tools used by software engineers to assist with code writing. Notable examples of such models integrated into development tools are GitHub Copilot[1], the world's most widely adopted AI developer tool, and CodeLlama [6], an open-source LLM dedicated to code generation.

These models are intended to work for everyone using any programming language [6, 9, 12]. However, they do not resemble similar performance for different programming languages [8, 10, 23]. In multilingual scenarios, the concept of low-resource programming languages emerges. A low-resource programming language is defined by two main criteria: 1) the programming language is either absent from the dataset or minimally represented in the corpus, and 2) there is limited availability of code samples online or low adoption of the language. This lack of representation and resources leads to reduced performance in these languages compared to more commonly used ones.

Consequently, the current state of research for estimating model performance across multiple languages cannot rely solely on popular evaluation metrics such as CodeBLEU score [20] or perplexity [23]. While these metrics are helpful for initial assessments of the ability to understand and replicate patterns, they are not sufficient indicators of functional correctness as they mainly evaluate based on similarities and patterns instead of real-world performance in producing code solutions. Therefore, they should not be proxied as comprehensive evaluations of the CLMs' capabilities in different languages.

Focusing on one of the most common downstream tasks, code generation, benchmarks are crucial for evaluating model performance. A benchmark is a standardized set of tasks or tests used to measure and compare the performance of different models. In the context of code generation, benchmarks often include extensive unit tests that provide quantifiable results based on the models' pass rate (i.e., the percentage of tests the model successfully completes). Research in this area has produced and adopted several key benchmarks such as HumanEval [9] and MBPP [2]. The main issue with these benchmarks is that they perpetuate the existing challenges of CLM performance in low-resource languages in the data corpus into the evaluation phase as most available benchmarks are limited to a single high-resource programming language, Python [2, 9, 13].

---

[1]https://github.com/features/copilot

This discrepancy in evaluation methodologies not being able to assess CLMs' performance in various languages has spurred substantial research efforts to translate the standardized benchmarks for evaluating the performance of CLMs into other programming languages. Therefore, this study aims to *address this issue by utilizing the latest advancements in translated code generation benchmarks, providing an initial assessment of CLMs in multi-lingual scenarios.* For that, we conducted a preliminary empirical study of two popular translated benchmarks, MultiPL-E [8] and HumanEval-SyntheSize [10], to evaluate the open-source CLM, PolyCoder's performance for code generation in them.

This research seeks to foster discussion and provide constructive feedback on the feasibility and potential of using these translated benchmarks as a novel and robust method for evaluating CLMs' performance. Our preliminary results offer promising insights, paving the way for further exploration and refinement in this cutting-edge area of software engineering.

The remainder of this paper is structured as follows: Section 2 covers the CLM, the translated benchmarks used in the study, and a detailed review of the related work. Section 3 details the methodology employed and the overall study design. The initial results are discussed in Section 4, and Section 5 goes over the implications of our initial results. Furthermore, Section 6 addresses the study's limitations and explains the future work. Finally, Section 7 concludes the paper.

## 2  BACKGROUND & RELATED WORK

This section will introduce several key concepts relevant to the current study. These include the translated benchmarks for evaluating code generation downstream tasks, the CLMs chosen for evaluation in this study, and an overview of previous related work in this area.

### 2.1  Code Generation Benchmarks

The code generation task is evaluated based on the code quality generated by CLMs. Various benchmarks are employed to assess the performance of CLMs in code generation tasks. Some benchmarks are automated and sourced from online platforms, such as the CoNala [24] benchmark, which uses StackOverflow data, and the Automated Programming Progress Standard (APPS) [7] benchmark, powered by coding platforms like Codewars. In contrast, there are manually curated datasets that aim to test functional correctness, such as HumanEval [9] and Mostly Basic Programming Problems (MBPP) [2]. Recent research in these manually curated benchmarks extends to finer details, such as class-level code generation [15].

Our study focuses on the HumanEval [9] benchmark released by the OpenAI team, which comprises 164 Python problems of varying difficulty. As a popular benchmark for evaluating CLMs in code generation, there have been several efforts to translate it into other languages. Notable examples include HumanEval-X [11], which provides datasets in four additional programming languages; Multi-HumanEval [5], which extends the dataset to eleven programming languages; and HumanEval-XL [19], which translates

**Table 1: Comparison of Characteristics of translated HumanEval Benchmarks**

| Characteristic | MultiPL-E | HumanEval Synthesize |
|---|---|---|
| Translated Languages | 18 | 5 |
| Methodology | Automated Tool | Manual |
| Extensible | Yes | No |
| Translation | Incomplete | Complete |

both the prompts and problems into 23 human languages and 12 programming languages.

Specifically, we choose the following two benchmarks for the ease of reproducibility and differences in their approach.

**MultiPL-E.** is a system specifically designed to translate code generation benchmarks from Python into multiple programming languages [8]. This automated tool utilizes a unique approach by employing language-specific transpilers, making it highly adaptable across various benchmarks, models, and languages.

**HumanEvalSynthesize.** This benchmark is a carefully curated translation of the HumanEval [9] benchmark into different programming languages. It is part of a larger suite of benchmarks called HumanEvalPack, which covers translated benchmarks for three software engineering downstream tasks, including code generation. This suite is introduced with the group of models, OctoPack [10], enhancing its benchmarking capabilities across multiple programming languages and three different software engineering downstream tasks.

The overall characteristics, as observed in the benchmark studies, are outlined in Table 1. MultiPL-E framework is more generic and can be extended to any code generation benchmark, such as MBPP [2]. While HumanEvalSynthesize is only focused on HumanEval code generation framework translation. Consequently, the latter can completely translate the whole problem set of the HumanEval framework into five programming languages. However, the former has to make some trade-offs to keep it more generalizable, covering up to 18 programming languages.

### 2.2  Code Language Model

Numerous CLMs have been developed to address software engineering tasks, with most production-ready models being closed-source, such as Codex [9], and a few smaller open-source models like CodeT5 [21]. However, recent studies have introduced open-source models pre-trained on multiple programming languages, such as CodeLlama [6]. In this study, we use PolyCoder [23], an open-source model designed to address the limitations of closed-source models. PolyCoder [23] is available in three different versions—160 million parameters, 0.4 billion parameters, and 2.7 billion parameters—trained on 12 different programming languages, including low-resource languages. We chose PolyCoder over CodeLlama because of its accessible sizes that align with the hardware we have. Additionally, the study [23] identified the problem of low-resource programming languages when they built the model and it wasn't already tested on MultiPL-E benchmark; instead, it used perplexity scores. This makes PolyCoder particularly suitable for our study, which aims to evaluate CLM performance across diverse languages using translated benchmarks.

## 2.3 Related work

Assessment of CLMs is a part of any study trying to introduce a model [12, 23], a benchmark [7, 9] or an advancement in existing work [16]. Recent advancements have seen a shift towards employing translated benchmarks to see the gap in research and capabilities of CLMs, notably the use of the MultiPL-E [8] approach in the CodeLlama paper [6]. Historically, before the availability of these sophisticated benchmarks, the multilingual performance of CLMs was gauged using metrics like perplexity, as demonstrated in the PolyCoder study [23]. This metric provided early insights but lacked the depth and specificity that translated benchmarks now offer. We intend to document this shift towards using newly available translated benchmarks.

ClassEval [15] empirically compares the existing state of different code generation benchmarks to see the gap in research and introduces their benchmark. This motivated us to study multilingual benchmarks, which lack a similar analysis. Due to the nascent research stage in translated code generation benchmarks, there are only acknowledgments of similar studies at the moment [5, 8, 10]. Still, these references to comparable research lack empirical and unbiased input. The insights gained in these studies are worth empirically verifying; for example, OctoCoder [10] translated benchmarks could see the best performance in the Python benchmark and the least in the low-resource programming language of Rust. Additionally, MultiPL-E [8] shows that perplexity scores measured by Xu et al. [23] do not co-relate to pass@1 in Codex. Hence, our study addresses this gap by initially evaluating multilingual benchmarks to enhance our understanding of their effectiveness and applicability.

## 3 STUDY DESIGN

This section provides an overview of the research questions addressed in this study and outlines the methodology employed to address them.

### 3.1 Research Question

This research is motivated by the need to understand how different translated benchmarks influence the performance of CLMs and to what extent these benchmarks can accurately reflect CLMs' capabilities in diverse linguistic environments. Thus, this study answers the following research question:

> **RQ:** How do translated benchmarks capture and reflect the multilingual performance of CLMs in both high-resource and low-resource programming languages?

This research question stems from an understanding that the efficacy of CLMs in multilingual contexts is crucial, given their expanding use across diverse programming environments. The question addresses the need to evaluate how varied methodological approaches within translated benchmarks that diverge from the original benchmark affect CLM's performance. As these models are applied to a broader array of languages, some of which may be underrepresented in training datasets, assessing whether the benchmarks used can accurately measure their capabilities and limitations is imperative. This inquiry aims to uncover potential disparities in performance across languages and identify reliable evaluation practices. By examining the consistency of CLM performance across different benchmarks, the study seeks to identify patterns and anomalies that could inform future development, refinement, and adoption of these tools. By exploring these aspects, the study aims to inform future enhancements in developing CLMs and the methodologies used to evaluate them, ensuring that the benchmarks are robust and comprehensive enough to reflect the data it's trained on.

### 3.2 Methodology

**Programming Languages Selection**: We considered four programming languages that are common in both of the translated benchmarks, including Python. JavaScript and Java were chosen because they are among the top four most widely used programming languages, and Rust was included as it is the top emerging language according to the GitHub October 2023 Survey[2]. Additionally, Java and JavaScript account for a high percentage in the final training data corpus (21.3% and 7.37%), while Rust represents one of the lowest (1.27%) among the twelve programming languages in the PolyCoder model [23]. This provides a good mix of multilingual and low-resource scenarios to answer our research question.

**Implementation of the CLMs:** The overview of the evaluation pipeline is illustrated in Figure 1. The process commenced with acquiring the three different sizes of a CLM: 160M, 0.4B, and 2.7B, as provided by PolyCoder [23] from the HuggingFace Hub using their Python library. Subsequently, the MultiPL-E benchmark was implemented by following the well-maintained tutorial provided by the original authors[3], utilizing an up-to-date GitHub repository for reference. Conversely, for the HumanEvalSynthesize benchmarks, we leveraged an actively maintained code generation framework introduced in the StarCoder model from the BigCode community [12]. As mentioned in Table 1, the MultiPL-E translated benchmarks are incomplete and had 161, 161, 158, and 156 problems in Python, JavaScript, Java, and Rust, respectively. On the other hand, the CLM was tested against the complete set of 164 questions in all translated benchmarks of HumanEvalSynthesize. All computations were executed on a high-performance computing offline node equipped with an NVIDIA Tesla V100 32GB NVLink GPU.

**Evaluation metrics:** The performance of the CLMs was evaluated using the pass@1 metric, with a conservative temperature setting of 0.2, as recommended by benchmarking studies [8, 10]. The pass@1 metric quantitatively measures the proportion of translated problems for which the CLM successfully passes all designated unit tests on the first attempt, providing a stringent test of the model's predictive accuracy in a controlled environment. Hence, the higher the pass@1 metric, the greater the ability to produce correct code solutions on its initial attempt for each given problem. The evaluation encompassed four programming languages: JavaScript, Java,

---

[2]https://github.blog/2023-11-08-the-state-of-open-source-and-ai/
[3]https://nuprl.github.io/MultiPL-E/tutorial.html

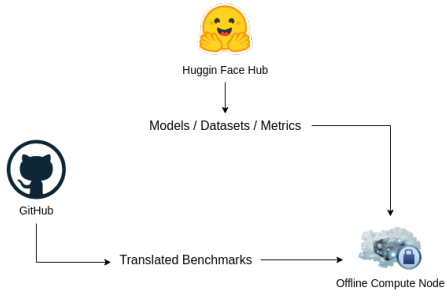

**Figure 1: Pipeline to evaluate CLMs in translated benchmarks**

Rust, and Python—the latter as originally defined in the HumanEval benchmark [9].

## 4 INITIAL RESULTS AND DISCUSSION

The pass rate of running the PolyCoder models against the MultiPL-E translated benchmarks in the first attempt (pass@1) can be seen in Table 2. Java HumanEval translation benchmark has the highest performance in larger models of 0.4B and 2.7B out of all languages. The next best and almost comparable result was observed in Python, followed by JavaScript, both considered high resource languages per the data corpus [23]. Finally, Rust had the lowest performance, which indicates its low resource nature.

The HumanEvalSyntheSize benchmark's pass@1 of PolyCoder models are illustrated in Table 3. Here, Python has the highest results across all models compared to the rest of the other languages. However, the second-best results for PolyCoder 2.7B, were seen in Rust, a low-resource programming language, followed by JavaScript and Java.

The evaluation involved assessing the models' performance across various sizes using the HumanEval Python version included in the replication packages of both benchmarks as discussed in Section 3.2. This setup was consistent with the pass@1 scores of 2.13%, 2.96 %, and 5.59% for 160M, 0.4B, and 2.7B size of PolyCoder respectively obtained in Xu et al. [23], ensuring the validity of the results as seen in Table 2 and 3. Interestingly, the highest performance apart from Python was observed in Java in MultiPL-E benchmarks, especially in larger sizes of the PolyCoder model (0.4B, 2.7B); this might be because of the larger Java size in the data corpus it was trained on and does not align with the perplexity scores the study relied on for testing multilingual nature [23]. Another observation from the superior performance in Java suggests that the translations and evaluation processes in the MultiPL-E approach may be more effective, providing a more accurate assessment of the models' capabilities. The performance of PolyCoder models is almost reversed in both translated benchmarks. In the MultiPL-E benchmark, Java performs best, as seen in Table 2, and Rust is the least performant. The opposite is observed in HumanEvalSynthe-size benchmarks as shown in Table 3. These differences could stem from the methodological differences and, most importantly, how the benchmarks are replicated. Implementing the HumanEvalSyn-thesize translation through the framework [3] might change the intended behavior of CLMs due to the trade-offs made during its integration to be compatible with Python libraries used while maintaining consistency across languages, which may drift apart from the paper [10]. Additionally, the performance in Rust was higher than in the MultiPL-E with a score of 3.05% in PolyCoder 2.7B compared to HumanEvalSynthesize, where it achieved a pass@1 rate of 2.30%. The differences in problem sizes, with 156 and 164 problems in the former and latter benchmarks respectively, could potentially account for the observed discrepancies in performance. It also suggests that HumanEvalSynthesize was able to create a better translation in Rust and capture the CLMs low-resource nature well.

A general pattern observed across both benchmarks was the improved performance with increased CLM size. Larger models consistently demonstrated better pass@1 rates in most languages. However, an exception to this trend was noted in Java within the HumanEvalSynthesize benchmark, where performance did not increase as expected with larger model sizes. This could again come from the implementation limitation of using the code generation framework [3]. Ultimately, the performance of CLMs is on the lower side, with the largest CLM of PolyCoder 2.7B giving 6.10%; this is still much lower when compared to the performance of state-of-the-art CLMs that are available out there [16], this is caused by the hardware limitation and not being able to accommodate larger CLMs. Consequently, such constraints magnify the impact of even minor inconsistencies in performance.

> **Finding** The CLMs exhibit inconsistent performance across the two translated benchmarks. However, a general pattern of improved performance is observed as the size increases and effectively captures the CLMs' performance across high-resource and low-resource languages.

**Table 2: Pass@1 scores for PolyCoder in MultiPL-E benchmarks**

| CLM | Python | JavaScript | Java | Rust |
|---|---|---|---|---|
| PolyCoder 160M | 2.73% | 1.52% | 1.71% | 0.80% |
| PolyCoder 0.4B | 3.31% | 3.18% | 3.83% | 1.88% |
| PolyCoder 2.7B | 5.56% | 5.16% | 5.63% | 2.30% |

**Table 3: Pass@1 scores for PolyCoder in HumanEvalSynthe-size benchmarks**

| CLM | Python | JavaScript | Java | Rust |
|---|---|---|---|---|
| PolyCoder 160M | 2.44% | 1.22% | 0.61% | 0.61% |
| PolyCoder 0.4B | 3.05% | 1.83% | 1.83% | 1.83% |
| PolyCoder 2.7B | 6.10% | 2.44% | 1.22% | 3.05% |

## 5 IMPLICATIONS

Considering the observed inconsistencies and performance variations across different benchmarks and languages detailed in Section 4, this section explores the implications of our study's findings and suggests avenues for improvement.

### 5.1 Implications for Researching Translated Benchmarks

The findings of this study suggest that translated versions of well-established benchmarks could be a valuable resource for evaluating the multilingual capabilities of CLMs. However, selecting these benchmarks should be preceded by a comprehensive evaluation of the methodological approaches employed in their development. This study highlighted that even aspects such as problem size can vary between benchmarks, leading to varying results. For instance, the inconsistencies observed between the two benchmarks in this study underscore the necessity of thorough validation.

Moreover, researchers should exercise caution when integrating new translated benchmarks into their studies. It is imperative to first verify the benchmark against existing models or studies to ascertain whether the replication yields consistent results. Such preliminary validations can help identify and mitigate potential discrepancies that might affect the robustness and reliability of research outcomes.

### 5.2 Implications for Developing Translated Benchmarks

The development of translated benchmarks is crucial for the progression of multilingual capabilities in CLMs. These benchmarks must be accompanied by comprehensive documentation and robust support mechanisms to maximize their effectiveness. Platforms like HuggingFace Hub provide an excellent infrastructure for distributing and utilizing machine learning models and datasets, and leveraging such platforms can significantly facilitate the adoption and replication of benchmarks across different research settings. Our study was constrained by the selection of benchmarks, which only included a limited number of common languages, with Rust being the sole representative of low-resource languages. This limitation is far from ideal and underscores the need for a broader approach to developing translated benchmarks. Expanding the range of languages covered by translated benchmarks would be immensely beneficial for the research community, particularly in enhancing model performance in underrepresented languages. Furthermore, there is significant potential in translating benchmarks to encompass domains beyond traditional coding problems. This suggestion is inspired by initiatives like [13], which propose extending the applicability and relevance of benchmarks to broader contexts. Finally, automating the translation process in benchmarks as shown in MultiPL-E [8] could potentially be the ideal way to extend the approach to various benchmarks, languages, and CLMs.

## 6 LIMITATIONS AND FUTURE WORK

This preliminary study presents foundational insights into the multilingual performance of CLMs using translated benchmarks while identifying several areas for future research and acknowledging inherent limitations.

The evaluation metrics used in this study primarily focused on pass rates at a threshold of 1. Expanding future evaluations to include pass rates at 10 and 20 could offer a more nuanced understanding of models' capabilities across varying difficulty levels. Additionally, integrating cutting-edge large language models, such as the Instruct variant of CodeLlama [4], could provide updated insights into the effectiveness of newer technologies in code generation tasks. A significant limitation noted in this study pertains to using the HumanEvalSynthesize benchmarks. The implementation of these benchmarks could be potentially limited by the tool used and methods employed by the BigCode community [3], which may differ from the original setup described in the OctoPack study [10].

In future work, we would like to explore parameter-efficient training methods [22] and quantized versions of larger CLMs to accommodate for insufficient resources or hardware to perform such studies. Moreover, extending the investigation to include other software engineering downstream tasks [16] would broaden the study's implications. Finally, analyzing the translations of various benchmarks offered by platforms such as CodeXGlue [14] could shed light on additional challenges and opportunities in the multilingual evaluation of CLMs.

## 7 CONCLUSION

In conclusion, this study has investigated the multilingual performance of CLMs by utilizing translated benchmarks. We introduce the concepts of code generation benchmarks and the need for translated versions to evaluate CLMs better. Our findings suggest that while translated benchmarks can serve as valuable tools for evaluating the capabilities of CLMs across different languages, their effectiveness is contingent upon the accuracy and consistency of the translation and benchmarking methodologies employed. Moreover, the notable variability between benchmarks emphasizes the necessity for researchers to undertake thorough validations and comparisons against existing models and literature to ensure the reliability and applicability of results. Looking forward, it is crucial to expand the scope of translated benchmarks to encompass a broader array of languages and domains, ultimately enhancing the generalizability of research findings and ensuring the robustness and applicability of CLMs across diverse real-world applications. This study contributes to the ongoing discourse in software engineering by shedding light on the challenges and opportunities associated with the multilingual evaluation of CLMs. By addressing these challenges, the research community can fully utilize translated benchmarks to improve the evaluation and development of CLMs.

## REFERENCES

[1] Toufique Ahmed and Premkumar Devanbu. 2023. Few-shot training LLMs for project-specific code-summarization. In *Proceedings of the 37th IEEE/ACM International Conference on Automated Software Engineering* (Rochester, MI, USA) *(ASE '22).* Association for Computing Machinery, New York, NY, USA, Article 177, 5 pages. https://doi.org/10.1145/3551349.3559555

[2] Jacob Austin, Augustus Odena, Maxwell Nye, Maarten Bosma, Henryk Michalewski, David Dohan, Ellen Jiang, Carrie Cai, Michael Terry, Quoc Le,

---

and Charles Sutton. 2021. Program Synthesis with Large Language Models. arXiv:2108.07732

[3] Loubna Ben Allal, Niklas Muennighoff, Logesh Kumar Umapathi, Ben Lipkin, and Leandro von Werra. 2022. A framework for the evaluation of code generation models. https://github.com/bigcode-project/bigcode-evaluation-harness.

[4] Antonio Mastropaolo et al. 2023. On the Robustness of Code Generation Techniques: An Empirical Study on GitHub Copilot. arXiv:2302.00438

[5] Ben Athiwaratkun et al. 2023. Multi-lingual Evaluation of Code Generation Models. arXiv:2210.14868

[6] Baptiste Rozière et al. 2024. Code Llama: Open Foundation Models for Code. arXiv:2308.12950

[7] Dan Hendrycks et al. 2021. Measuring Coding Challenge Competence With APPS. arXiv:2105.09938

[8] Federico Cassano et al. 2022. MultiPL-E: A Scalable and Extensible Approach to Benchmarking Neural Code Generation. arXiv:2208.08227 [cs.LG]

[9] Mark Chen et al. 2021. Evaluating Large Language Models Trained on Code. arXiv:2107.03374 [cs.LG]

[10] Niklas Muennighoff et al. 2024. OctoPack: Instruction Tuning Code Large Language Models. arXiv:2308.07124

[11] Qinkai Zheng et al. 2023. CodeGeeX: A Pre-Trained Model for Code Generation with Multilingual Evaluations on HumanEval-X. arXiv:2303.17568

[12] Raymond Li et al. 2023. StarCoder: may the source be with you! arXiv:2305.06161

[13] Rongao Li et al. 2023. TACO: Topics in Algorithmic COde generation dataset. *arXiv preprint arXiv:2312.14852* (2023).

[14] Shuai Lu et al. 2021. CodeXGLUE: A Machine Learning Benchmark Dataset for Code Understanding and Generation. *CoRR* abs/2102.04664 (2021).

[15] Xueying Du et al. 2023. ClassEval: A Manually-Crafted Benchmark for Evaluating LLMs on Class-level Code Generation. arXiv:2308.01861

[16] Xinyi Hou et al. 2024. Large Language Models for Software Engineering: A Systematic Literature Review. arXiv:2308.10620

[17] Mohamad Khajezade, Jie JW Wu, Fatemeh Hendijani Fard, Gema Rodriguez-Perez, and Mohamed Sami Shehata. 2024. Investigating the Efficacy of Large Language

[18] Rangeet et al. Pan. 2024. Lost in Translation: A Study of Bugs Introduced by Large Language Models while Translating Code. In *Proceedings of the IEEE/ACM 46th International Conference on Software Engineering (ICSE '24)*. ACM. https://doi.org/10.1145/3597503.3639226

[19] Qiwei Peng, Yekun Chai, and Xuhong Li. 2024. HumanEval-XL: A Multilingual Code Generation Benchmark for Cross-lingual Natural Language Generalization. In *Proceedings of the 2024 Joint International Conference on Computational Linguistics, Language Resources and Evaluation (LREC-COLING 2024)*, Nicoletta Calzolari, Min-Yen Kan, Veronique Hoste, Alessandro Lenci, Sakriani Sakti, and Nianwen Xue (Eds.). ELRA and ICCL, Torino, Italia, 8383–8394. https://aclanthology.org/2024.lrec-main.735

[20] Shuo Ren, Daya Guo, Shuai Lu, Long Zhou, Shujie Liu, Duyu Tang, Neel Sundaresan, Ming Zhou, Ambrosio Blanco, and Shuai Ma. 2020. CodeBLEU: a Method for Automatic Evaluation of Code Synthesis. arXiv:2009.10297

[21] Yue Wang, Weishi Wang, Shafiq Joty, and Steven C. H. Hoi. 2021. CodeT5: Identifier-aware Unified Pre-trained Encoder-Decoder Models for Code Understanding and Generation. arXiv:2109.00859

[22] Martin Weyssow, Xin Zhou, Kisub Kim, David Lo, and Houari Sahraoui. 2024. Exploring Parameter-Efficient Fine-Tuning Techniques for Code Generation with Large Language Models. arXiv:2308.10462

[23] Frank F. Xu, Uri Alon, Graham Neubig, and Vincent J. Hellendoorn. 2022. A Systematic Evaluation of Large Language Models of Code. arXiv:2202.13169 [cs.PL]

[24] Pengcheng Yin, Bowen Deng, Edgar Chen, Bogdan Vasilescu, and Graham Neubig. 2018. Learning to Mine Aligned Code and Natural Language Pairs from Stack Overflow. In *International Conference on Mining Software Repositories (MSR)*. ACM, 476–486. https://doi.org/10.1145/3196398.3196408

Models for Code Clone Detection. In *Proceedings of the 32nd IEEE/ACM International Conference on Program Comprehension* (<conf-loc>, <city>Lisbon</city>, <country>Portugal</country>, </conf-loc>) *(ICPC '24)*. Association for Computing Machinery, New York, NY, USA, 161–165. https://doi.org/10.1145/3643916.3645030